# Ptychographic reconstruction algorithm for frequency resolved optical gating: super-resolution and supreme robustness

**PAVEL SIDORENKO,\* OREN LAHAV, ZOHAR AVNAT AND OREN COHEN\***

*Department of Physics and Solid State Institute, Technion, Haifa 32000, Israel*
*\*Corresponding authors: oren@technion.ac.il, sidor@tx.technion.ac.il*

**Frequency-resolved optical gating (FROG) is probably the most popular technique for complete characterization of ultrashort laser pulses. In FROG, a reconstruction algorithm retrieves the pulse amplitude and phase from a measured spectrogram, yet current FROG reconstruction algorithms require and exhibit several restricting features that weaken FROG performances. For example, the delay step must correspond to the spectral bandwidth measured with large enough SNR – a condition that limits the temporal resolution of the reconstructed pulse, obscures measurements of weak broadband pulses, and makes measurements of broadband mid-IR pulses hard and slow because the spectrograms become huge. We develop a new approach for FROG reconstruction, based on ptychography (a scanning coherent diffraction imaging technique), that removes many of the algorithmic restrictions. The ptychographic reconstruction algorithm is significantly faster and more robust to noise than current FROG algorithms, that are based on generalized projections (GP). We demonstrate, numerically and experimentally, that ptychographic reconstruction works well with very partial spectrograms, e. g. spectrograms with reduced number of measured delays and spectrograms that have been substantially spectrally filtered. In addition, we implement the ptychogrpahic approach to blind second harmonic generation (SHG) FROG and demonstrate robust and complete characterization of two unknown pulses from a single measured spectrogram and the power spectrum of only one of the pulses. We believe that the ptychograpy-based approach will become the standard reconstruction procedure in FROG and related diagnostics methods, allowing successful reconstructions from so far unreconstructable spectrograms. © 2016 Optical Society of America**

## 1. INTRODUCTION

Frequency-resolved optical gating (FROG) is probably the most commonly used method for full characterization (i.e. amplitude and phase) of ultrashort optical pulses [1,2]. A FROG apparatus produces a two-dimensional (2D) intensity diagram (spectrogram), also known as FROG trace, of an input pulse by interacting the pulse with its delayed replica in a nonlinear-optical medium, e.g. second harmonic generation (SHG) crystal [3]. Current FROG reconstruction procedures [4–6] are based on 2D phase retrieval algorithms [7,8], somewhat similar to the approach used in 2D coherent diffraction imaging [9]. These generalized projections (GP) algorithms [4–6] require Fourier relation between the frequency and delay axes of the measured spectrogram: $\Delta\omega\Delta t = 1/N$ where $\Delta t$ is the delay step, $\Delta\omega$ is the spectral resolution and N is the number of delay steps as well as the number of frequencies measured for each delay. The reconstruction resolution of the GP algorithms is limited by the spectrogram delay step which must corresponds to the spectral bandwidth of the trace [5]. This algorithmically imposed coupling between the reconstruction resolution, delay step and measured spectral bandwidth with large enough SNR weakens FROG performances in several ways. For example, it dictates that the bandwidths of the nonlinear medium and the spectrometer must be large enough in order to not filter the spectrogram. Indeed, this is the reason why crystals of SHG FROG for measuring ultrashort pulses are very thin: such crystals support broad bandwidth phase matching. Unfortunately, the price for using thin crystals is reduced conversion efficiency and SNR. This feature limits the application of FROG in measurement of weak broad-bandwidth laser pulses. A hardware based method was suggested to effectively increase the phase-matching bandwidth of SHG FROG [10], but it requires mechanical scanning of the SHG crystal orientation, which significantly increases the measurement acquisition time, and still provides limited bandwidth improvement. Another problem that results from the imposed Fourier relation is that spectrograms of ultrashort mid-IR pulses are often huge, resulting with computationally very slow reconstruction algorithms [11]. Also, an algorithm that will work with partial spectrograms (i.e. spectrograms that do not conform to the Fourier relation) may allow to significantly reduce the number of scanning steps in FROG apparatus, yielding much faster measurements. Finally, there is always motivation to increase the robustness of FROG reconstruction algorithms to noise.

Reconstruction algorithms that work well even with partial spectrograms were recently developed for several characterization techniques of ultrashort laser pulses [12–16], but not for FROG. Especially prominent is the recent adaptation of the general principle of ptychography – a scanning coherent diffraction imaging method [17] – to diagnostics of femtosecond pulses using cross-correlation FROG (XFROG) [14,16] and attosecond pulses from FROG-CRAB (Frequency-Resolved Optical Gating for Complete Reconstruction of Attosecond

Bursts) measurements [13]. These works demonstrated the superb robustness of the ptychographic-based reconstruction approach, both in terms of SNR and the use of only partial spectrograms. Specifically, reconstruction from reduced number of delay steps was demonstrated. However, the temporal resolution of the recovered field was still limited in these works by the measured spectral bandwidth (reconstruction with spectral filtering was not considered). Overall, the implementation of the ptychographic-based approach to pulse diagnostic techniques (XFROG and FROG-CRAB) in which the unknown pulse interacts with another pulse that is fully or partially known was shown to be very successful. However, so far this approach has not been adapted to techniques like FROG, in which the unknown pulse interacts with its exact replica and therefore the reconstruction problem is more difficult.

Here we propose and demonstrate experimentally a ptychography-based pulse reconstruction algorithm for FROG that does not require any prior information on the pulse. We show that the proposed algorithm outperforms the conventional GP FROG pulse recovery method in terms of noise-robustness. Moreover, we also demonstrate that our algorithm can successfully recover pulses from much less measurements than are needed in current pulse recovery algorithms, which are based on GP principle. We demonstrate numerically and experimentally recovery of a pulse even when the delay or frequency axis are under-sampled considerably or truncated by a low pass filter. We also demonstrate that if additional information about the pulse is known in advance, e.g. its power spectrum, then successful pulse recovery is possible even from ridiculously small number of measurements. Finally, we explore the application of the ptychographic approach to blind FROG, where two unknown pulses are characterized in parallel. We found that the two pulses can be completely and reliably characterized from a single spectrogram and the power spectrum of one of the pulses.

## 2. RECONSTRUCTION PROCEDURE

Our method is based on ptychography, which is a very powerful CDI technique that has recently gained a remarkable momentum in optical microscopy in the visible, extreme ultraviolet and x-ray spectral region [17]. In ptychography, a complex-valued object is scanned in a step-wise fashion through a localized coherent beam. In each scanning step, the intensity of the diffraction pattern of the object is measured, typically in a Fraunhofer plane. The diffraction pattern is associated with Fourier transform of the illuminated part of the object. Critically, the spatial support of the illuminating spot must be bigger than the step size so that neighboring diffraction patterns will result from different, but overlapping, regions of the object. The set of measured diffraction patterns is used for reconstructing the complex valued transmittance of the object. Ptychography exhibits several advantages over "conventional" imaging phase retrieval schemes, including significant improvement in the robustness to noise, no requirement for prior information (e.g. support) on the object and probe beam, no loss of information due to the beam stops, and generally faster and more reliable reconstruction algorithms [18]. Moreover, the uniqueness in one [19] and two [20] dimensional ptychography is guaranteed, provided that the illuminating beam is known. Ptychography has been proven to be useful in case where the illumination beam is known [18,20], but also when the illumination beam is unknown. In the latter case, the unknown illumination is reconstructed together with the sought object [21,22].

We first describe our ptychography-based pulse recovery algorithm for FROG (specifically SHG-FROG) in details. A SHG-FROG trace [3] is given by

$$I_{FROG}^{SHG}(\omega, \Delta t) = |\int_{-\infty}^{\infty} E(t)E(t-\Delta t)e^{-i\omega t}dt|^2 \quad (1)$$

where $E(t)$ is the complex envelope of the unknown pulse (the carrier frequency has been removed) and $\Delta t$ is the delay step between the pulse and its replica. A convenient discrete form of Eq. (1) for pulse recovery is:

$$I_j(\omega) = |F[\chi_j(t)]|^2 \quad (2)$$

where $\chi_j(t) = E(t)E(t-j\Delta t)$, , $j = 1 \ldots J$ is a running step delay index, $\Delta t$ is the delay step and $F$ stands for Fourier transform operator. Our iterative recovery algorithm starts with initial guess for the unknown pulse, $E_o(t)$. In our case, it corresponds to the integrated measured FROG trace over the frequency axis (this initial guess corresponds to the intensity autocorrelation of the pulse [23]). Each iteration of the recovery algorithm starts with producing a new array of the integers $1, 2 \ldots J$, yet in a random order, $s(j)$. This array is used for randomizing the order of step delays. Within each iteration, the recovered pulse is modified J times using an internal loop (with running index j). We now describe the sequence of steps to obtain the j-th modification within the internal loop, i.e. the steps to yield the updated recovered field $E_{j+1}(t)$, from the field $E_j(t)$, and the measured spectrogram at time delay $s(j)$, $I_{s(j)}(\omega)$. First, the SHG signal of the field is calculated:

$$\psi_j(t) = E_j(t)E_j(t-s(j)\Delta t) \quad (3)$$

Second, the SHG signal is Fourier transformed and its modulus is replaced by the square-root of the $s(j)$-ordered measured spectrum. Importantly, we replace only the part that was measured reliably. Other spectral components are not changed ($\Omega$ marks the set of reliably measured spectral components

$$\Phi_j^{\Omega}(\omega) = \sqrt{I_{S(j)}^{\Omega}(\omega)} \frac{F[\psi_j(t)]}{|F[\psi_j(t)]|} \quad (4)$$

In addition, we suggest applying a soft thresholding procedure (a well-known procedure for solving linear inverse problems [24]) for the subset of unknown frequencies ($\Omega^C$ i.e. $\omega \not\in \Omega$), for both the real and imaginary parts of the signal by:

$$\Phi_j^{\Omega^C}(\omega) = f_\gamma^{ST}(\text{Re}\{F[\psi_j(t)]\}) + i f_\gamma^{ST}(\text{Im}\{F[\psi_j(t)]\}) \quad (5)$$

where the soft threshold operation $f_\gamma^{ST}(x)$ is defined by

$$f_\gamma^{ST}(x) = \begin{cases} 0 & \text{if } x < \gamma \\ x - \gamma \text{sign}(x) & \text{if } x \geq \gamma \end{cases} \quad (6)$$

where γ is a threshold parameter, typically set between $10^{-3}$ and $10^{-6}$. This optional procedure improves the reconstruction results. Intuitively, one can consider the soft thresholding as a de-noising procedure [25,26] which is applied selectively on the unknown spectral part of the trace. Third, updated SHG signal is calculated by:

$$\psi'_j(t) = F^{-1}[\Phi_j(\omega)] \quad (7)$$

Finally, the pulse is updated with a weight function based on the complex conjugate of $E_j^*(t)$ according to:

$$E_{j+1}(t) = E_j(t) + \alpha \frac{E_j^*(t-s(j)\Delta t)}{|E_j(t-s(j)\Delta t)|^2_{max}} (\psi'_j(t) - \psi_j(t)) \quad (8)$$

In Eq. (8), $\alpha$ is a real parameter that controls the strength of the update. Crucially, in our algorithm a new $\alpha$ is selected randomly in each iteration (its probability is distributed uniformly in the range $[0.1, 0.5]$)). The iterations continue until the stopping criteria (difference between measured and calculated FROG traces is smaller than SNR) is reached.

Our algorithm is based on the extended ptychographical iterative engine (ePIE) [22] that is commonly used in ptychography. We adopted it for FROG through three modifications. First, we replaced the illuminating probe beam by the delayed replica of the sought pulse, i.e. by $E(t-j\Delta t)$. Second, we introduced randomly varying $\alpha$. We found that convergence with fixed $\alpha$ requires a very large number of runs of the algorithm with random initial guesses (typically>100) while with randomly varying $\alpha$, we found out that ~95% of the random initial guesses converge. Intuitively, randomization of $\alpha$ significantly reduces the probability of stagnation in a local minimum (the advantages of using random $\alpha's$ in phase retrieval algorithms are discussed in Ref. [27]). A third optional modification is the replacement of only part of the power spectrum in the 2nd step of the algorithm, i.e. Eq. 4, with or without the soft thresholding procedure. This modification allows us to reconstruct pulses at high resolution from incomplete spectrograms (as shown below in Fig. 2). More modifications are presented below: to allow utilization of additional measured information (power spectrum) and modification to ptychographic blind FROG reconstruction.

## 3. RESULTS

### A. Numerical results with complete spectrograms

Next, we characterize the performances of our ptychographic-based FROG algorithm as a function of SNR, and compare it with a commonly used GP-based reconstruction FROG algorithm: Principal Component Generalized Projections Algorithm (PCGPA) [28,29]. To this end, we numerically produced a set of 100 laser pulses that all conform to a Gaussian power spectrum that is centered at 800nm and its spectral bandwidth is 213nm. Each pulse (N=128 grid points) is produced by applying a random spectral chirp to the above power spectrum, (conditioned that the support of the pulse is ≤200 fs). The calculated FROG traces are 128 by 128 points with equally spaced delays, $\Delta t = 1.57$ fs, and spanning the same frequency window (i.e. the product of the delay and spectral resolutions is $\frac{1}{N} = \frac{1}{128}$). White-Gaussian noise σ is added to the simulated FROG trace, at different SNR values, defined by: $SNR = 20\log(\|I_{FROG}\|/\|\sigma\|)$, where $\|\cdot\|$ stands for $l_2$ norm. **Fig.**

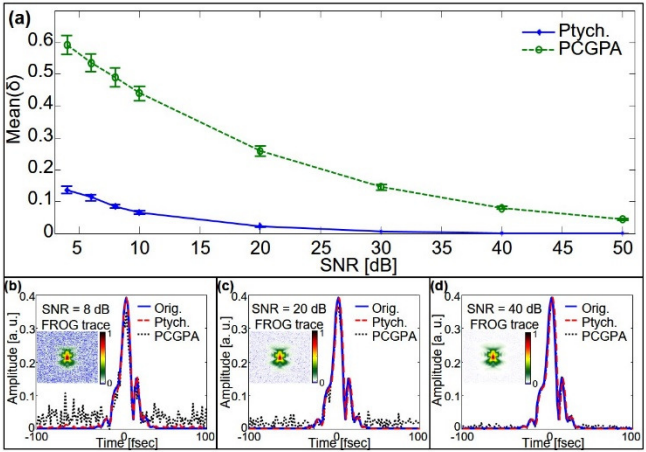

**1.** (a) Mean angle (representing the error) between reconstructed and true signals as a function of SNR for ptychographic-based algorithm (solid blue curve) and for PCGPA method (dashed green curve). Error bars represent standard deviation of the average over 100 pulses. Plots (b), (c) and (d) show a true pulse (blue solid curves), reconstructed by ptychographic-based algorithm (red dotted curves) and by PCGPA method (dashed black curve) for SNR=8dB, 20dB and 40dB respectively.

Next, we recovered the set of 100 unknown pulses using both ptychographic-based and PCGPA algorithms, each one with 10 independent runs (the initial pulse in PCGPA was random) where each run is limited to 1000 iterations and chose the best solution according to the stopping criteria (minimal NMSE between simulated and recovered traces). In order to characterize the quality of the reconstructions, one needs to take into account that SHG FROG suffers from the following ambiguities: trivial time shift, time direction, and global phase. Thus, we first removed the above first two ambiguities from the reconstructed fields and then evaluate the error of the reconstructions using the following angle

$$\delta(E, \hat{E}) = \arccos\left(\frac{|\langle \hat{E}(t)|E(t)\rangle|}{\sqrt{\langle \hat{E}(t)|\hat{E}(t)\rangle\langle E(t)|E(t)\rangle}}\right) \quad (9)$$

which is insensitive to the global phase (i.e. the 3rd ambiguity) of the field [27]. In Eq. 9, $\hat{E}(t)$ and $E(t)$ are the original and reconstructed fields, respectively, and $\langle f(t)|g(t)\rangle = \int f(t)\overline{g(t)}dt$ stands for inner product. Figure 1a shows the mean angle between the reconstructed and original pulses as a function of SNR for ptychographic-based algorithm (solid blue curve) and PCGPA method (dashed green curve). Figure 1b, 1c and 1d display recovery of a representative pulse with SNR=8dB, SNR=20dB, SNR=40dB respectively. Clearly, the ptychographic-based reconstruction algorithm outperforms PCGPA significantly, for all the explored SNR values.

### B. Numerical results with incomplete spectrograms

In order to apply PCGPA and all other current FROG reconstruction methods, the measured FROG trace should consist of N×N (i.e. N spectral at N time delays) measurements and the product of the delay step and spectral resolution should be 1/N [23]. Also, the bandwidth of the FROG trace should be ~1.4 times larger than the bandwidth of the pulse power spectrum autocorrelation [23]. We term such FROG traces as complete traces. As discussed in the introduction section, these constrains weaken the performances of FROG and impose hardware limitations in the FROG apparatus. However, the source for these requirements is algorithmic and not fundamental. After all, complete FROG traces are highly redundant: there are many informative measurements in the trace while the pulse is a vector of only N complex numbers. In addition, the FROG trace depends nonlinearly on the pulse, e.g. it mixes its spectral components. Indeed, we will show below that the ptychographic reconstruction approach can reconstruct pulses from significantly incomplete FROG traces. We define an incompleteness parameter by:

$$\eta = \frac{\text{\# of pixels in the incomplete trace}}{\text{\# of pixels in the complete trace}}$$

Particularly, we will present super-resolution in FROG, where the complete FROG trace and its associated pulse are retrieved from a FROG trace which was spectrally low-pass filtered (such filter can result from the limited bandwidth of the nonlinear medium or the spectrometer [23,30]). Equivalently, to super-resolution in imaging, bandwidth extrapolation in FROG corresponds to temporal super resolution. That is, q times bandwidth extrapolation corresponds to q times super-resolution. The ptychographic approach can reconstruct pulses from their incomplete FROG traces because in each updating step of the reconstructing process, it makes use of a single measured spectrum (measured at one specific delay value). As a result, the delay and spectral axes are completely uncoupled. Moreover, the delay step grid does not define the temporal grid (and therefore also the temporal resolution) of the reconstructed pulse. In addition, the delay step does not need to be constant. Moreover, as shown below, the resolution of the reconstructed pulse can be higher than one may expect from the measured spectral bandwidth and Fourier uncertainty. Figure 2 presents typical results of such reconstructions using a pulse from the bank of pulses used for Fig. 1. (An example of a pulse with very high time bandwidth product (TBP) is presented in Fig. S1 in the supplementary information section). Simulated FROG traces of the pulse, without and with 20 dB noise are presented in Figs. 2a and 2b, respectively. In Figs 2(c-v), each horizontal panel presents results for a different type of truncation. Importantly, in all the reconstructions presented in this figure, only the incomplete (truncated) noisy FROG traces were fed into the reconstruction algorithm. No prior information about the pulses was assumed. Figs. 2c-g show reconstructions from incomplete FROG traces that were spectrally truncated by a low pass filter (LPF). Figure 2c shows a FROG trace obtained by filtering the trace in Fig. 2b by a window-shape filter that is centered at $\omega=0$ which conserves the information in only 13 (out of 128) frequencies and nullifies all the other frequencies ($\eta \cong 0.1$). Applying the ptychographic reconstruction on the FROG trace in Fig. 2c yields the reconstructed (extrapolated) FROG trace in Fig. 2d and the associated reconstructed pulse with amplitude and phase that are shown in Figs. 2e and 2f, respectively, compared with the original pulse. The calculated angle (error) of this reconstruction is $\delta \cong 0.04$. Figure 2g shows the angle between the reconstructed and original pulses as a function of the incompleteness parameter, clearly showing that the reconstruction works well until the incompleteness parameter approaches $\eta = 0.1$. In the second horizontal panel, the frequency axis was under sampled. Figure 2h shows a FROG trace where 7/8 of the spectral lines of the complete noisy trace in Fig. 2b were nullified ($\eta \cong 0.125$). Applying the ptychographic reconstruction on the FROG trace in Fig. 2h yields the reconstructed FROG trace in Fig. 2i and the associated reconstructed pulse with amplitude and phase that are shown in Figs. 2j and 2k, respectively ($\delta \cong 0.055$). Figure 2l shows that the reconstruction is good until the incompleteness parameter approaches $\eta \cong 0.125$. We applied the soft thresholding option (Eqs. 5 and 6) with $\gamma = 5 \times 10^{-6}$ in the reconstructions presented in the first two panels of Fig. 2. Reconstructions without applying soft thresholding are presented in Fig. 2S in the supplementary information section. The 3rd horizontal panel presents reconstruction from FROG traces that were truncated by a temporal window-shape filter that is centered at $\Delta t=0$. Figure 2m shows a FROG trace with 22 delay points ($\eta=0.172$). The reconstructed trace is shown in Fig. 2n. The original and reconstructed temporal amplitude and phase are shown in Figs. 2o and 2p, respectively ($\delta \cong 0.2$). Figure 2q shows the reconstruction error as a function of the incompleteness parameter. The bottom panel presents reconstructions from spectrograms with under-sampled delay axis. Fig. 2r shows a FROG trace in which only 13 (out of 128) equally-spaced delay points were conserved ($\eta \approx 0.1$). The reconstructed FROG trace is shown in Fig. 2s. The original and reconstructed temporal amplitude and phase are shown in Figs. 2t and 2u, respectively ($\delta \cong 0.04$). Figure 2v shows that the reconstruction is good until the incompleteness parameter approaches $\eta \cong 0.1$.

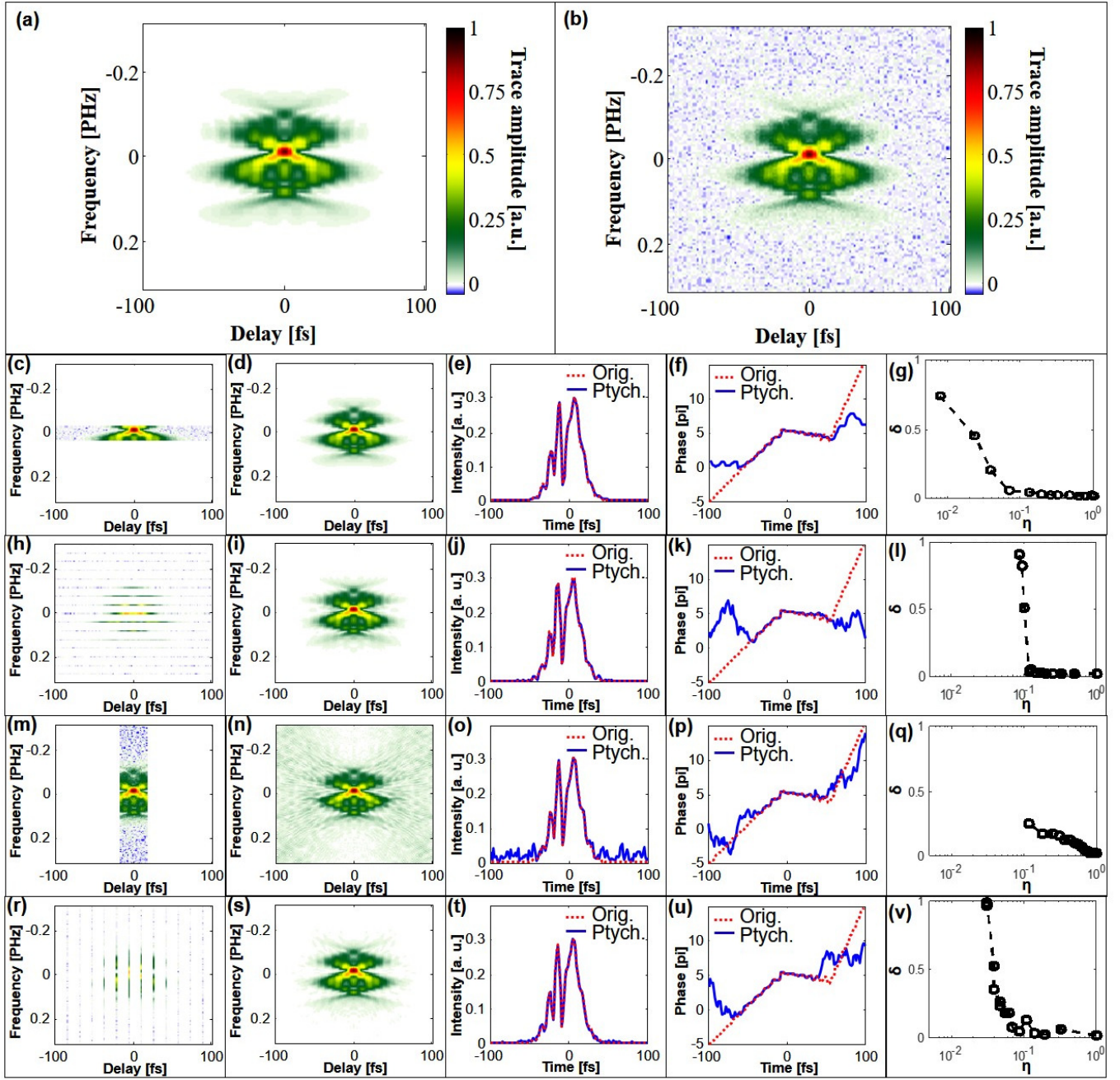

**Fig. 2.** Examples of pulse reconstructions from incomplete FROG traces using ptychographic-based reconstruction algorithm. Numerically simulated complete FROG traces without (a) and with 20 dB noise (b). All reconstructions below are from incomplete traces obtained by truncating the noisy trace in plot b. Top horizontal panel (Figs. 2c-g) presents reconstruction from frequency filtered traces. Second horizontal panel (Figs. 2h-l) presents reconstructions from spectrally under sampled traces. Third panel (Figs. 2m-q) present reconstructions from temporally filtered traces. Last bottom panel (Figs. 2r-v) presents reconstructions from temporally under sampled trace. In each panel, the 1st (left) and 2nd (left) plots show an example of the incomplete FROG trace and its respective reconstructed trace. The 3rd and 4th plots present the amplitude and phase of the original and reconstructed pulses, respectively. The 5th plot shows the angle between the reconstructed and original pulses (i.e. the reconstruction error) as a function of the incompleteness parameter. All partial and reconstructed traces are presented with same colormap as in plot (a).

### C. Ptychographic FROG: Experimental results

We demonstrate our ptychographic-based FROG algorithm experimentally, using our home-made SHG-FROG. The laser pulse was produced by an ultrafast Ti-sapphire laser system. The measurement was done with 3 fs delay step and 512 delay points. Hence, the complete FROG trace consists of 512×512 data points (Fig. 3a). We applied the ptychographic-based and PCGPA algorithms and obtained the reconstructed FROG traces in Fig. 3b and 3c: the NMSE between the measured and reconstructed traces are 0.0351 and 0.0363, respectively. Figures 3d and 3e present the reconstructed amplitude and phase by the ptychographic-based algorithm (blue solid curve) and the PCGPA algorithm (red dashed curve) respectively. The good correspondences between the reconstructions demonstrate that the ptychographic based algorithm works well also in experiments. Next, we demonstrate experimentally the capability of ptychographic-based FROG reconstruction to retrieve the pulse from incomplete FROG traces. In this case, PCGPA and other conventional GP based FROG reconstruction algorithms fail. Thus, in Fig. 4 we compare the reconstruction from the incomplete traces with the ptychographic reconstruction using the full spectrogram in Fig. 3. The top horizontal panel in Fig. 4 presents reconstruction from a spectrally low pass filtered FROG trace. The used trace with 128 nonzero spectral lines ($\eta \cong 0.25$) is shown in Fig. 4a. The second horizontal panel shows pulse reconstruction from spectrally under-sampled FROG trace. Figure 4e presents the used trace where 75% of the spectral lines were nullified ($\eta \cong 0.25$). The third horizontal panel presents reconstruction from temporally filtered trace with 210 delays ($\eta \cong 0.41$). Finally, the forth panel presents reconstruction from temporally undersampled spectrogram: only each 8th delay step was conserved ($\eta \cong 0.125$) (Fig. 4m). In each panel, the second plot from the left presents the reconstructed FROG traces (for comparison see measured trace in Fig. 3a) from the corresponding incomplete measured trace to its left. The third and the fourth columns show the reconstructed amplitude and phase of the pulse from full (red dashed curve) and corresponding incomplete measured (blue solid curve) traces, respectively. The reconstructions match quite well (note that the deviations in the reconstructed phases are significant only in the region of low amplitudes), demonstrating experimentally the capability of the ptychographic-based reconstruction algorithm to reconstruct the pulses from incomplete spectrograms.

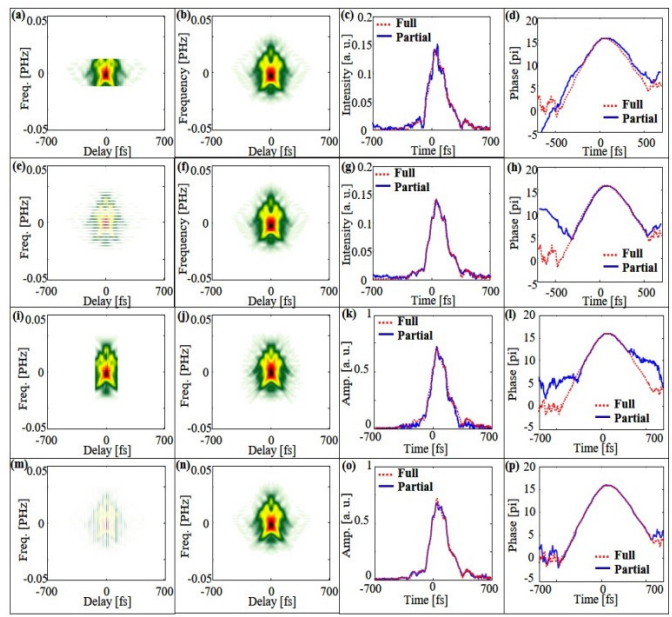

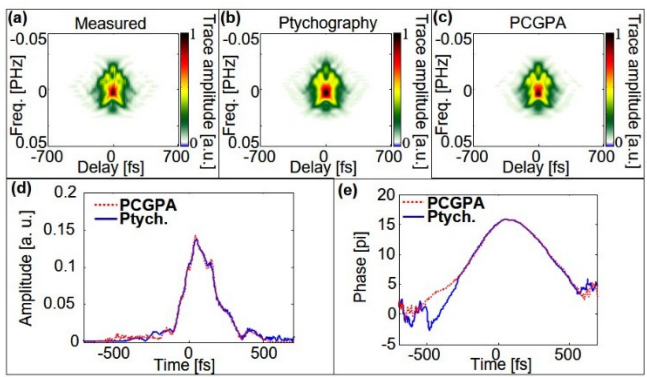

**Fig. 3.** Demonstration of ptychographic-based reconstruction using experimental SHG-FROG data. (a) measured FROG trace. (b) and (c) are traces recovered by ptychgraphic-based algorithm and PCGPA algorithm respectively. Amplitude (d) and phase (e) of the recovered pulse by ptychographic-based algorithm (solid blue curve) and PCGPA algorithm (red dashed curve) respectively.

**Fig. 4.** Experimental pulse reconstructions from incomplete FROG traces with. First horizontal panel (a-d) presents reconstruction from low pass filtered trace. Second horizontal panel (e-h) presents reconstruction from spectrally under-sampled trace. Third horizontal panel (i-l) presents reconstruction from a trace that was low pass filtered in the delay axis. Forth horizontal panel (m-p) presents reconstruction from under sampled trace in the delay axis. In each panel, the second plot from the left presents the reconstructed FROG traces from the corresponding incomplete measured trace to its left. The second and the third columns show the reconstructed amplitude and phase of the pulse from full (red dashed curve) and corresponding incomplete measured (blue solid curve) traces, respectively. All partial and reconstructed traces are presented with same colormap as in Fig 3a.

### D. Ptychographic FROG with measured power spectrum

The power spectrum of the measured pulse is often available, or it can be measured by a spectrometer. In current FROG reconstruction algorithms, the directly measured power spectrum may be used for estimating the consistency of the experimental results and the quality of the reconstruction [23]. However, it is not utilized within the reconstruction procedure. Adding the directly measured spectrum as additional constrain in the reconstruction process [31] leads to instabilities due to the required deconvolution, hence it is not useful [3]. In our ptychographic-based approach, on the other hand, addition of the directly measured spectrum does not lead to instability because we do not perform deconvolution and, as shown above, we can use only part of the FROG trace (the part with high SNR).

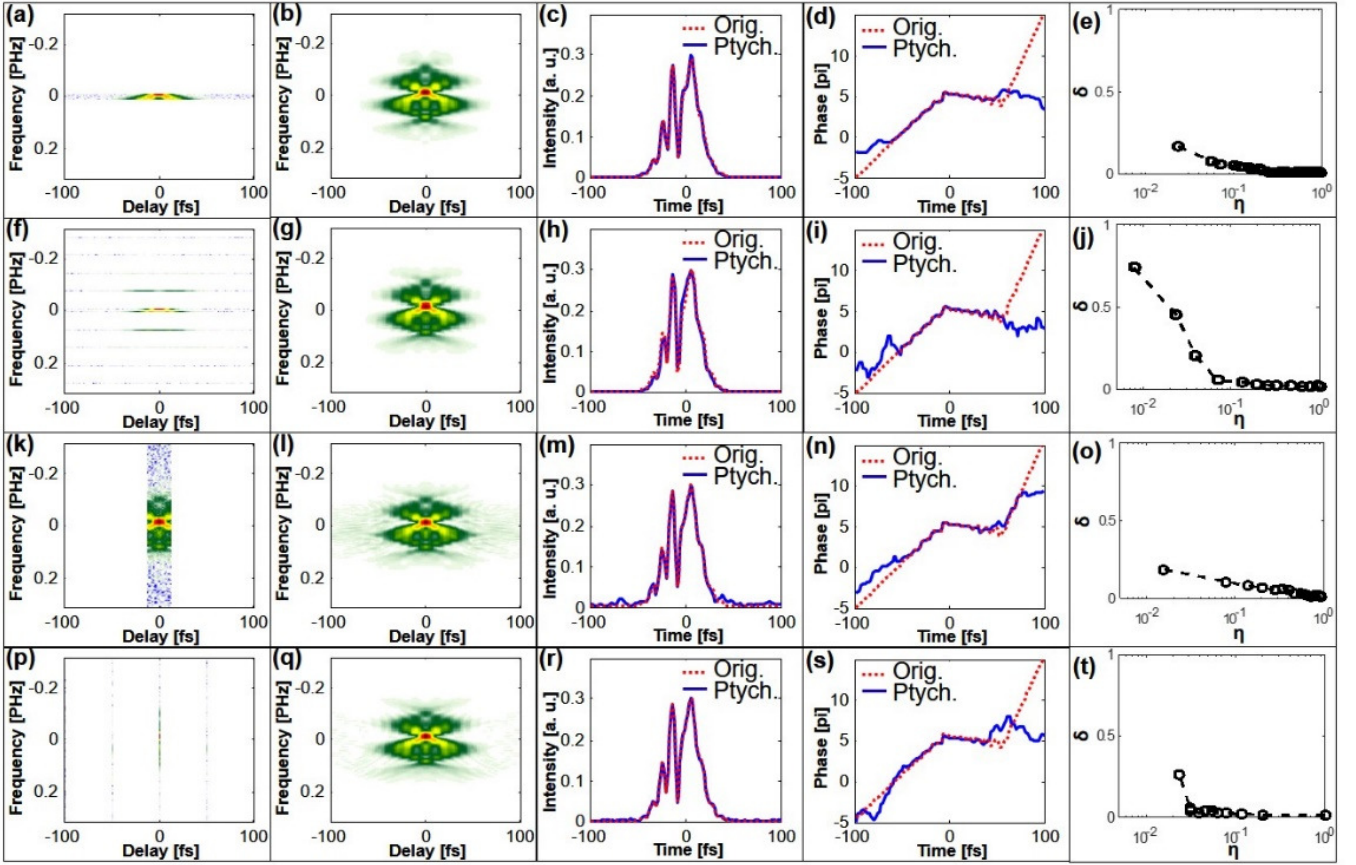

**Fig. 5.** Pulse reconstructions from incomplete FROG traces and pulse power spectrum. We use the same exemplary pulse that was used in Fig. 2. The structure of Figs. 5(a-t) is the same as the structure of Figs. 2(c-v). All partial and reconstructed traces are presented with same colormap as in Fig. 2a.

We employ the measured power spectrum to our ptychographic-based reconstruction scheme by projecting the updated pulse in each iteration to the sub-space with the given power spectrum. We add a fifth step in each iteration to the algorithm described in the method section (the fourth step corresponds to Eq. 8). We first replace the power spectrum of the updated pulse by the directly measured power spectrum

$$\widehat{E}_{j+1}(\omega) = \sqrt{\hat{I}(\omega)} \frac{F[E_{j+1}(t)]}{|F[E_{j+1}(t)]|} \quad (10)$$

where $\hat{I}(\omega)$ is the measured pulse power spectrum. Next we obtain the time-domain reconstructed pulse at the end of the j+1 iteration.

$$\widehat{E}_{j+1}(t) = F^{-1}[\widehat{E}_{j+1}(\omega)] \quad (11)$$

We have found that the ptychographic-based FROG algorithm with a power spectrum constrain works very well, even when the FROG trace is significantly spectrally filtered or under sampled. Numerical results using the same exemplary pulse that was used in Fig. 2 are displayed in Fig. 5. The structure of Fig. 5 is the same as the structure of Figs. 2(c-v). As in Fig. 2, all the reconstructions presented in Fig. 5 use FROG traces that are filtered from the trace shown in Fig. 2b. The first horizontal panel in Fig. 5 shows pulse reconstruction from a spectrally low pass filtered trace that conserved only 5 frequencies ($\eta \cong 0.04$, $\delta \cong 0.05$). The second panel presents pulse reconstruction from spectrally under-sampled trace ($\eta \cong 0.08$, $\delta \cong 0.09$). Third and fourth panels present reconstructions from temporally filtered ($\eta \cong 0.125$, $\delta \cong 0.08$) and temporally under-sampled ($\eta \cong 0.03$, $\delta \cong 0.04$) traces respectively. Comparing the right columns in Figs. 2 and 5 clearly show the strength of adding the measured power spectrum to the reconstruction algorithm.

### E. Incomplete spectrograms

We already presented several examples of pulse reconstructions from incomplete FROG traces, both theoretically (Figs. 2 and 5) and experimentally (Fig. 4). Recalling that GP-based algorithms require complete spectrograms, it is natural to ask whether the opportunity for reconstruction from incomplete spectrograms is general or is it limited to several examples. Thus, we tested our algorithm on many hundreds of different pulses and found that all the pulses could be reconstructed from significantly incomplete FROG traces. Figure 6 presents results from such a test. In this case, we first numerically produced a set of 100 very different pulses (N=128 grid points) by the following procedure. To get a random power spectrum, we Fourier transformed a complex function with a Gaussian-shaped amplitude with FWHM = 150fs and a pseudo-random phase (obtained by applying a moving average window of 5 points to a random vector) and then take an absolute value of the obtained complex function. Next, we multiply the obtained power spectrum by a pseudo-random spectral phase. Finally, we Fourier transform the pulse (back to time domain) and include it in the set if its support is ≤200 fs. We repeated this procedure until obtaining 100 pulses in the set. Next, we calculated the complete FROG traces of the pulses and added 20db white Gaussian noise. We then applied the ptychographic reconstructed algorithm on the filtered noisy FROG traces. In Fig. 6 we present results for spectral low pass filter and delay under sampling that in our opinion are the most important for applications. Figures 6a and 6b present the mean angle of reconstruction as a function of incompleteness parameter for frequency filtered traces and for delay under sampling, respectively, with spectral prior information (red solid curve) and without any prior information (blue dashed curve). From our experience, $\delta \leq 0.1$ correspond to very good reconstructions (Figs. 6c-v show four reconstruction examples with $\delta \sim 0.1$). Clearly, good reconstructions were obtained even when the FROG traces were significantly filtered, especially if the power spectrum is included as prior information.

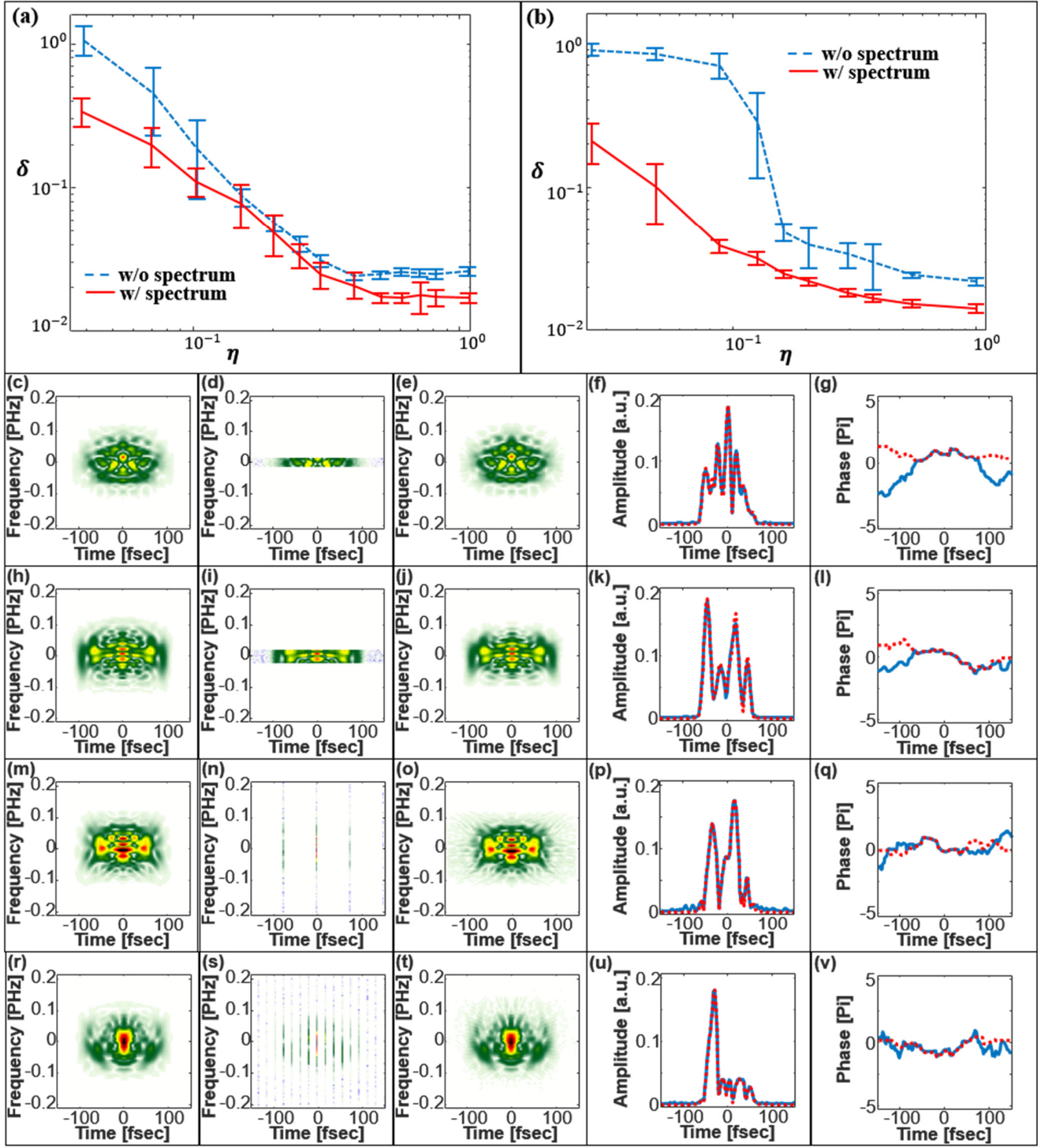

**Fig. 6.** Statistical investigation of pulse reconstructions from incomplete FROG traces. Reconstruction mean angle as a function of incompleteness parameter for frequency filtered traces (a) and for delay under sampling (b) with spectral prior information (red solid curve) and without any prior information (blue dashed curve), error bars corresponds to standard deviation over 100 pulses. Four exemplary pulse reconstructions: Spectrogram is spectrally low pass filtered, with (plots (c-g), $\eta$=0.07, $\delta$=0.096] and without [plots(h-l), $\eta$=0.102, $\delta$=0.093] power spectrum prior information. Spectrogram is delay under sampled, with [plots(m-q), $\eta$=0.031, $\delta$=0.093] and without [plots(r-v), $\eta$=0.125, $\delta$=0.085] power spectrum prior information. Each panel shows the complete FROG traces (left column plots), filtered trace, reconstructed trace, original (dashed red) and reconstructed (solid blue) pulse amplitudes and pulse phases in the right column plots. All partial and reconstructed traces are presented with same colormap as in Fig. 2a.

### F. Ptychographic blind FROG

FROG setups can typically measure spectrograms produced by two different pulses. Indeed, in some experiments, one would like to fully characterize two potentially different ultrashort laser pulses simultaneously [32–34]. This scenario is known as blind FROG [2,35]. Assuming SHG FROG, the measured spectrogram in this case is given by:

$$I_{\text{blind FROG}}^{\text{SHG}}(\omega, \Delta t) = |\int_{-\infty}^{\infty} P(t)G(t-\Delta t)e^{-i\omega t}dt|^2 \quad (12)$$

where P(t) and G(t) are the complex envelopes of the two pulses, that without losing generality we term probe and gate pulses. Retrieving the two pulses from a blind FROG trace was proved to be ill-posed, i.e. many different pulse pairs yield the same measured trace [36,37] (notably, non-trivial ambiguities are practically absent in blind polarization gating (PG) FROG, yet PG-FROG is insensitive to the phase of one of the pulses [35], hence it does not lead to complete characterization of two different pulses). It was shown that the ambiguity was removed if the power spectra of the two pulses were also measured (provided that the two spectra are different) and measured FROG trace is non-centrosymmetric [36]. However, even in this case no reconstruction algorithm was proposed. Thus, blind FROG is generally not useful, unless additional information, like additional trace with a different gate pulse is available [38]. As shown below, we adopted the ptychographic-based FROG reconstruction algorithm to blind FROG. We found that it works efficiently and robustly. Furthermore, we discovered that single spectrogram and power spectrum of one of the pulses (the gate) is enough information for retrieving both pulses.

In this paragraph, we present the list of modifications in the ptychographic-based FROG algorithm (presented in section 2) in order to apply it to blind FROG case. First, the algorithm starts with the following initial pulses: the gate pulse is transform limited (with the measured power spectrum) and the probe pulse is zero. Second, Eq. 1 is replaced by Eq. 12. The discrete form of Eq. 2 is not changed. Third, the SHG signal is defined now by: $\chi_j(t) = P(t)G(t-j\Delta t)$, hence Eq. 3 is replaced by Eq. 13:

$$\psi_j(t) = P_j(t)G_j(t-s(j)\Delta t) \quad (13)$$

Fourth, Eq. 8 is replaced by Eqs. 14 and 15, where both pulses are now updated in each iteration:

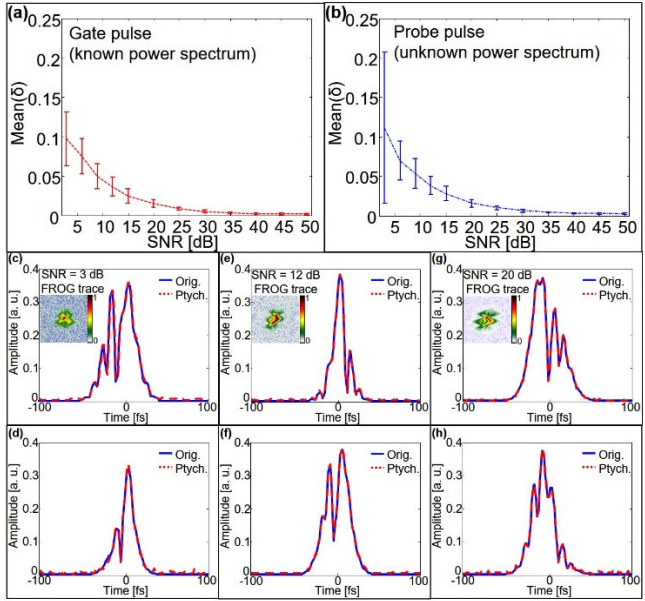

**Fig. 7.** Numerical characterization of the ptychographic blind FROG reconstruction algorithm. Mean angle (representing the error) between reconstructed and original gate (a) and probe (b) pulses as a function of SNR. Error bars represent standard deviation of the average over 100 pairs. Reconstruction examples are shown in plots c-h. The upper and lower panels show original (blue solid curves) and reconstructed (red dotted curves) gate and probe pulses, respectively. The SNR is 3dB in plots (c) and (d), 12dB in plots (e) and (f) and 20dB in plots (g) and (h).

$$P_{j+1}(t) = P_j(t) + \alpha \frac{G_j^*(t-s(j)\Delta t)}{|G_j(t-s(j)\Delta t)|_{\max}^2}(\psi_j'(t) - \psi_j(t)) \quad (14)$$

$$G_{j+1}(t-s(j)\Delta t) = G_j(t) + \alpha \frac{P_j^*(t-s(j)\Delta t)}{|P_j(t-s(j)\Delta t)|_{\max}^2}(\psi_j'(t) - \psi_j(t)) \quad (15)$$

Finally, additional (fifth) step is added to each iteration of the algorithm. The gated pulse is projected to its measured power spectrum:

$$\hat{G}_{j+1}(\omega) = \sqrt{\hat{I}_G(\omega)} \frac{F[G_{j+1}(t)]}{|F[G_{j+1}(t)]|} \quad (16)$$

where $\hat{I}_G(\omega)$ is the measured power spectrum of the gate pulse. The time-domain reconstructed gate at the end of the j+1 iteration is given by:

$$\hat{G}_{j+1}(t) = F^{-1}[\hat{G}_{j+1}(\omega)] \quad (17)$$

Next, we characterize the performances of our ptychographic-based blind FROG algorithm as a function of SNR. We randomly chose 100 pairs of pulses from the set of 100 pulses that was described at the beginning of section 3.A and was used for calculating Fig. 1. White-Gaussian noise σ was added to the simulated blind FROG trace at different SNR values.

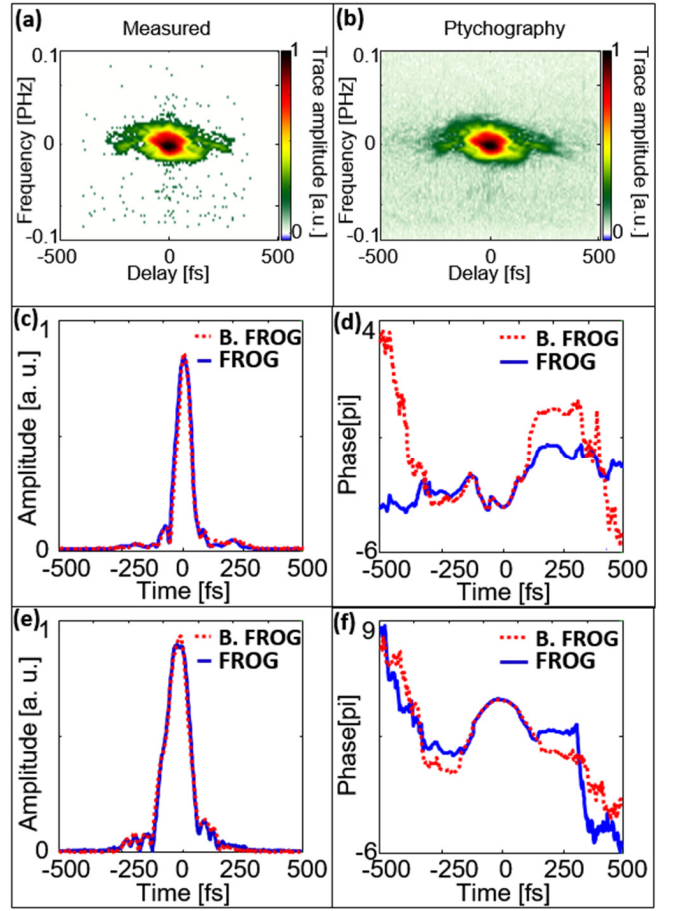

**Fig. 8.** (a) Measured blind FROG trace. (b) Trace recovered by ptychographic-based blind FROG algorithm (NMSE=0.092). (c) Amplitude (c) and phase (d) of the first pulse recovered from blind FROG (red doted curve) and FROG (solid blue curve). Amplitude (e) and phase (f) of the second pulse recovered from blind FROG (red doted curve) and FROG (solid blue curve).

Next, we apply the ptychographic-based blind FROG algorithm to reconstruct the unknown probe and gate pulses from the noisy trace and the power spectrum of the gate pulse. Figures 7a and 7b show the mean angle (average over 100 pairs) between the reconstructed and original gate and probe pulses, respectively, as a function of SNR. Figures 7c-h display three examples of recovery of probe and gate pulses with different SNR values. Clearly, the ptychographic-based blind FROG reconstruction algorithm correctly recovered all pairs of probe-gate pulses up to a noise level defined by SNR.

Finally, we present experimental blind FROG pulse reconstruction using our ptychographic-based reconstruction algorithm. One laser pulse was produced directly from a ti:sapphire laser-amplifier system. The second pulse was produced by passing a replica of the first pulse through 5mm of glass. We measured a blind FROG spectrogram and the power spectrum of the pulse which passed through the glass. For reference, we also measured each pulse separately using FROG. The FROG and blind FROG measurements were done with 2 fs delay and 512 delay points, i.e. each trace consists of 512×512 data points. The measured blind FROG trace is clearly asymmetric (Fig. 8a). We applied the ptychographic-based blind FROG algorithm and obtained the reconstructed trace in Fig. 8b (NMSE between the measured and reconstructed traces is 0.092). Figures 8c and 8d show the amplitude and phase, respectively, of the first pulse recovered from the blind FROG (red doted curve) and FROG (solid blue curve) traces. Angle between those two recovered pulses is $\delta = 0.11$. Figures 8e and 8f show the amplitude and phase, respectively, of the second pulse recovered from the blind FROG (red doted curve) and FROG (solid blue curve) traces. Angle between those two recovered pulses is $\delta = 0.15$. The good correspondences between blind FROG and FROG reconstructions demonstrate that the ptychographic based blind FROG algorithm works well also in experiments.

## 4. CONCLUSIONS

In conclusions, we proposed, characterized and demonstrated, numerically and experimentally, ptychographic-based algorithm for FROG trace inversion. Ptychographic-based algorithm allows retrieving the complex pulse from a very small number of delay steps and/or from a spectrally incomplete FROG trace. In addition, we applied the ptychographic approach to blind FROG and discovered that we can robustly recover the two pulses from a single measured blind FROG trace and power spectrum of one of the pulses. These new algorithmic capabilities will surely open many new opportunities in diagnostics of ultrashort laser pulses. The fact that our procedure can successfully recover pulses from significantly spectrally filtered spectrograms should allow measuring ultrashort laser pulses with resolution that is much higher than the corresponding bandwidth of the nonlinear process. For example, this feature can be useful for measuring weak pulses using thick nonlinear crystals that on one hand yield improved SNR but on the other hand introduce phase matching spectral filtering. Remarkably, in FROG CRAB, it may allow measurement of zeptosecond temporal structures [39,40] where finding a nonlinear process with the required spectral bandwidth is challenging. Notably, the ptychographic-based algorithm can significantly upgrade all FROG devices without any hardware modification, leading to successful reconstructions of pulses that were so far unmeasurable with those devices.

Finally, it is worth noting two directions that may further improve and extend the scope of ptychographic pulse reconstruction algorithms: 1) we expect that utilizing the oversampling in the spectral axes can be useful for better reconstructions (the spectral resolution of most current FROG devices greatly exceeds the measured temporal window, yet this additional information is lost because current reconstruction of GP based algorithms require Fourier relation between the spectral and delay axes). 2) An exciting direction is to utilize structure-based prior knowledge on the laser pulses in order to further enhance the resolution and noise robustness [41–43].

We invite researchers to download (freely) our ptychographic SHG FROG from our website:
http://tx.technion.ac.il/~oren/pty_FROG.html

## REFERENCES

1. R. Trebino, K. W. DeLong, D. N. Fittinghoff, J. N. Sweetser, M. A. Krumbügel, B. A. Richman, and D. J. Kane, "Measuring ultrashort laser pulses in the time-frequency domain using frequency-resolved optical gating," Rev. Sci. Instrum. **68**, 3277 (1997).
2. R. Trebino, *Frequency-Resolved Optical Gating: The Measurement of Ultrashort Laser Pulses* (Springer US, 2000).
3. K. W. DeLong, R. Trebino, J. Hunter, and W. E. White, "Frequency-resolved optical gating with the use of second-harmonic generation," J. Opt. Soc. Am. B **11**, 2206 (1994).
4. J. R. Fienup, "Reconstruction of a complex-valued object from the modulus of its Fourier transform using a support constraint," J. Opt. Soc. Am. A **4**, 118 (1987).
5. R. Trebino and D. J. Kane, "Using phase retrieval to measure the intensity and phase of ultrashort pulses: frequency-resolved optical gating," J. Opt. Soc. Am. A **10**, 1101 (1993).
6. R. P. Millane, "Multidimensional phase problems," J. Opt. Soc. Am. A **13**, 725 (1996).
7. J. R. Fienup, "Phase retrieval algorithms: a comparison.," Appl. Opt. **21**, 2758–69 (1982).
8. J. R. Fienup, "Phase retrieval algorithms: a personal tour [Invited].," Appl. Opt. **52**, 45–56 (2013).
9. Y. Shechtman, Y. C. Eldar, O. Cohen, H. N. Chapman, J. Miao, and M. Segev, "Phase Retrieval with Application to Optical Imaging: A contemporary overview," IEEE Signal Process. Mag. **32**, 87–109 (2015).
10. P. O'Shea, M. Kimmel, X. Gu, and R. Trebino, "Increased-bandwidth in ultrashort-pulse measurement using an angle-dithered nonlinear-optical crystal," Opt. Express **7**, 342 (2000).
11. P. K. Bates, O. Chalus, and J. Biegert, "Ultrashort pulse characterization in the mid-infrared.," Opt. Lett. **35**, 1377–9 (2010).
12. M. Chini, S. Gilbertson, S. D. Khan, and Z. Chang, "Characterizing ultrabroadband attosecond lasers.," Opt. Express **18**, 13006–16 (2010).
13. M. Lucchini, M. H. Brügmann, A. Ludwig, L. Gallmann, U. Keller, and T. Feurer, "Ptychographic reconstruction of attosecond pulses," Opt. Express **23**, 29502 (2015).
14. D. Spangenberg, E. Rohwer, M. H. Brügmann, and T. Feurer, "Ptychographic ultrafast pulse reconstruction," Opt. Lett. **40**, 1002 (2015).
15. J. Gagnon, E. Goulielmakis, and V. S. Yakovlev, "The accurate FROG characterization of attosecond pulses from streaking measurements," Appl. Phys. B **92**, 25–32 (2008).
16. D. Spangenberg, P. Neethling, E. Rohwer, M. H. Brügmann, and T. Feurer, "Time-domain ptychography," Phys. Rev. A **91**, 021803 (2015).
17. J. M. Rodenburg, "Ptychography and related diffractive imaging methods," Adv. Imaging Electron Phys. **150**, 87–184 (2008).
18. J. M. Rodenburg and H. M. L. Faulkner, "A phase retrieval algorithm for shifting illumination," Appl. Phys. Lett. **85**, 4795 (2004).
19. M. Guizar-Sicairos, K. Evans-Lutterodt, A. F. Isakovic, A. Stein, J. B. Warren, A. R. Sandy, S. Narayanan, and J. R. Fienup, "One-dimensional hard x-ray field retrieval using a moveable structure," Opt. Express **18**, 18374 (2010).
20. B. . McCallum and J. . Rodenburg, "Two-dimensional demonstration of Wigner phase-retrieval microscopy in the STEM configuration," Ultramicroscopy **45**, 371–380 (1992).
21. P. Thibault, M. Dierolf, O. Bunk, A. Menzel, and F. Pfeiffer, "Probe retrieval in ptychographic coherent diffractive imaging.," Ultramicroscopy **109**, 338–43 (2009).
22. A. M. Maiden and J. M. Rodenburg, "An improved ptychographical phase retrieval algorithm for diffractive imaging," Ultramicroscopy **109**, 1256–1262 (2009).
23. K. W. DeLong, D. N. Fittinghoff, and R. Trebino, "Practical issues in ultrashort-laser-pulse measurement using frequency-resolved optical gating," IEEE J. Quantum Electron. **32**, 1253–1264 (1996).
24. A. Beck and M. Teboulle, "A Fast Iterative Shrinkage-Thresholding Algorithm for Linear Inverse Problems," SIAM J. Imaging Sci. **2**, 183–202 (2009).
25. D. L. Donoho, "De-noising by soft-thresholding," IEEE Trans. Inf.

Theory **41**, 613–627 (1995).
26. A. Buades, B. Coll, and J. M. Morel, "A Review of Image Denoising Algorithms, with a New One," Multiscale Model. Simul. **4**, 490–530 (2005).
27. M. Köhl, A. A. Minkevich, and T. Baumbach, "Improved success rate and stability for phase retrieval by including randomized overrelaxation in the hybrid input output algorithm," Opt. Express **20**, 17093 (2012).
28. D. J. Kane, "Recent progress toward real-time measurement of ultrashort laser pulses," IEEE J. Quantum Electron. **35**, 421–431 (1999).
29. D. J. Kane, "Principal components generalized projections: a review [Invited]," J. Opt. Soc. Am. B **25**, A120 (2008).
30. B. Yellampalle, K. Kim, and A. J. Taylor, "Amplitude ambiguities in second-harmonic generation frequency-resolved optical gating," Opt. Lett. **32**, 3558 (2007).
31. J. Paye, M. Ramaswamy, J. G. Fujimoto, and E. P. Ippen, "Measurement of the amplitude and phase of ultrashort light pulses from spectrally resolved autocorrelation," Opt. Lett. **18**, 1946 (1993).
32. V. G. Lyssenko, J. Erland, I. Balslev, K.-H. Pantke, B. S. Razbirin, and J. M. Hvam, "Nature of nonlinear four-wave-mixing beats in semiconductors," Phys. Rev. B **48**, 5720–5723 (1993).
33. M.-A. Mycek, S. Weiss, J.-Y. Bigot, S. Schmitt-Rink, D. S. Chemla, and W. Schaefer, "Femtosecond time-resolved free-induction decay of room-temperature excitons in GaAs quantum wells," Appl. Phys. Lett. **60**, 2666 (1992).
34. A. Freiberg and P. Saari, "Picosecond spectrochronography," IEEE J. Quantum Electron. **19**, 622–630 (1983).
35. K. W. DeLong, R. Trebino, and W. E. White, "Simultaneous recovery of two ultrashort laser pulses from a single spectrogram," J. Opt. Soc. Am. B **12**, 2463 (1995).
36. B. Seifert, H. Stolz, and M. Tasche, "Nontrivial ambiguities for blind frequency-resolved optical gating and the problem of uniqueness," J. Opt. Soc. Am. B **21**, 1089 (2004).
37. D. J. Kane, G. Rodriguez, A. J. Taylor, and T. S. Clement, "Simultaneous measurement of two ultrashort laser pulses from a single spectrogram in a single shot," J. Opt. Soc. Am. B **14**, 935 (1997).
38. A. Consoli, V. Chauhan, J. Cohen, L. Xu, P. Vaughan, F. J. Lopéz-Hernández, and R. Trebino, "Retrieving Two Pulses Simultaneously and Robustly Using Double-Blind FROG," in *Frontiers in Optics 2010/Laser Science XXVI* (OSA, 2010), p. FThD8.
39. G. Mourou and T. Tajima, "Physics. More intense, shorter pulses.," Science **331**, 41–2 (2011).
40. C. Hernández-García, J. A. Pérez-Hernández, T. Popmintchev, M. M. Murnane, H. C. Kapteyn, A. Jaron-Becker, A. Becker, and L. Plaja, "Zeptosecond high harmonic keV x-ray waveforms driven by midinfrared laser pulses.," Phys. Rev. Lett. **111**, 033002 (2013).
41. S. Gazit, A. Szameit, Y. C. Eldar, and M. Segev, "Super-resolution and reconstruction of sparse sub-wavelength images," Opt. Express **17**, 23920 (2009).
42. A. Szameit, Y. Shechtman, E. Osherovich, E. Bullkich, P. Sidorenko, H. Dana, S. Steiner, E. B. Kley, S. Gazit, T. Cohen-Hyams, S. Shoham, M. Zibulevsky, I. Yavneh, Y. C. Eldar, O. Cohen, and M. Segev, "Sparsity-based single-shot subwavelength coherent diffractive imaging," Nat. Mater. **11**, 455–459 (2012).
43. P. Sidorenko, O. Kfir, Y. Shechtman, A. Fleischer, Y. C. Eldar, M. Segev, and O. Cohen, "Sparsity-based super-resolved coherent diffraction imaging of one-dimensional objects," Nat. Commun. **6**, 8209 (2015).

# Supplementary Information

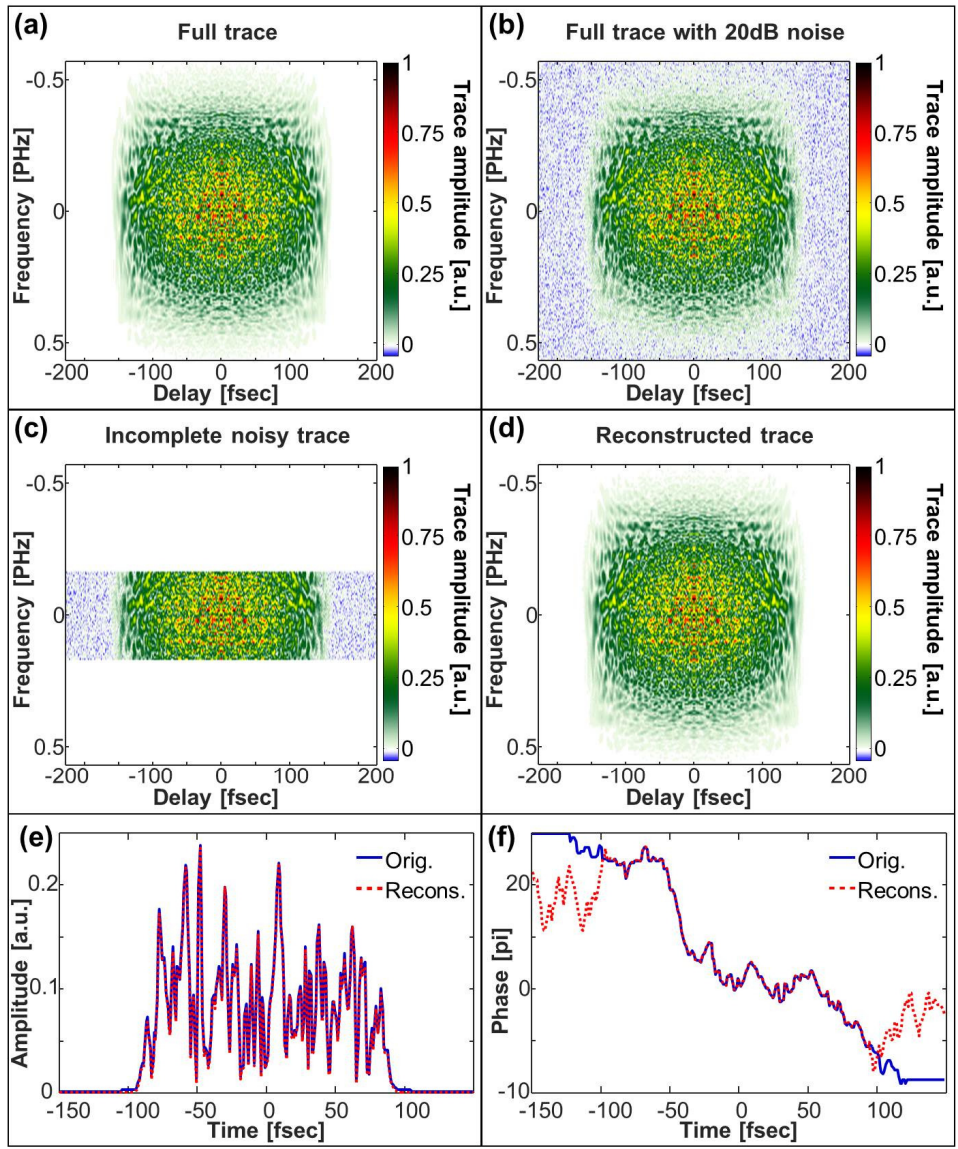

**Fig. 1S**. Numerical example of a complicated pulse (Time-Bandwidth Product $\cong$ 100) recovery from spectrally filtered trace. Numerically simulated complete FROG traces without (a) and with 20 dB noise (b). (c) Noisy and spectrally filtered trace ($\eta \cong 0.3$) fed into the ptychographic reconstruction algorithm (without any prior information). (d) Recovered trace. Original (blue solid curve) and reconstructed (red dashed curve) temporal amplitude (e) and phase (f) respectively. Angle between original and reconstructed pulses is $\delta \cong 0.016$

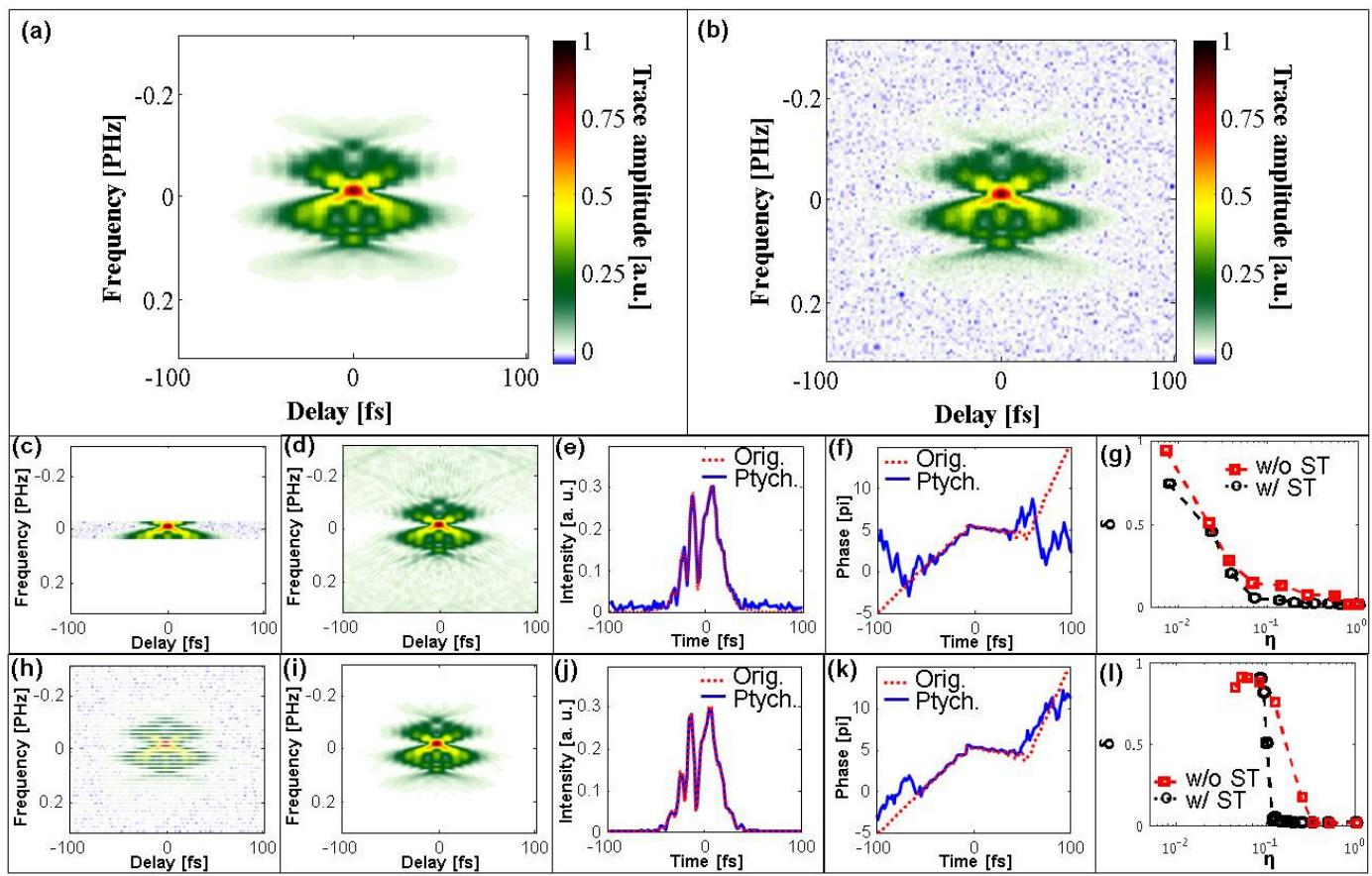

**Fig. 2S.** Pulse reconstructions from incomplete FROG traces using ptychographic-based reconstruction algorithm without applying the soft thresholding option (compare with Fig. 2 where the soft thresholding option was applied). Figs 2S(a) and 2S(b) are identical to Figs. 2a and 2b, displaying numerically simulated complete FROG traces without (a) and with 20 dB noise (b). Top horizontal panel (Figs. 2S c-g) presents reconstruction from frequency filtered traces ($\eta \cong 0.1$, $\delta \cong 0.138$). Second horizontal panel (Figs. 2S h-l) presents reconstructions from spectrally under sampled traces ($\eta \cong 0.33$, $\delta \cong 0.035$). In each panel, the 1st (left) and 2nd (left) plots show an example of the incomplete FROG trace and its respective reconstructed trace. The 3rd and 4th plots present the amplitude and phase of the original and reconstructed pulses, respectively. The 5th plot shows the angle between the reconstructed and original pulses (i.e. the reconstruction error) as a function of the incompleteness parameter with (black circles) and without (red squares) soft thresholding option. All partial and reconstructed traces are presented with same colormap as in plot (a).